# Rupture sismique des fondations par perte de capacité portante: Le cas des semelles circulaires

Foundation seismic bearing capacity failure: the case study of circular footings

C. T. Chatzigogos, A. Pecker, J. Salençon

*Laboratoire de Mécanique des Solides, CNRS UMR 7649, Département de Mécanique, École Polytechnique, France*

RÉSUMÉ

Cette étude, placée dans le cadre de la conception parasismique de fondations superficielles, concerne la détermination de la capacité portante sismique d'une semelle circulaire qui repose à la surface d'un sol semi infini purement cohérent hétérogène. Dans la première partie de l'article, une base de données contenant des cas d'ouvrages et de structures de génie civil ayant subi une rupture par perte de capacité portante au niveau de la fondation, est présentée, contribuant ainsi à la compréhension du phénomène étudié. En parallèle, les cas d'études contenus dans la base de données peuvent être utiles pour la validation de calculs théoriques. Dans la seconde partie de l'article, le problème susmentionné est traité par l'approche cinématique du Calcul à la Rupture et des bornes supérieures optimales sont établies pour les charges sismiques ultimes supportées par le système sol-fondation. Les résultats conduisent à une méthodologie simple qui permet l'utilisation des formules analytiques proposées dans l'Eurocode 8 (valables pour la portance sismique des semelles filantes sur sols homogènes) dans le cas des semelles circulaires et dans celui des sols cohérents hétérogènes.

ABSTRACT

Within the context of earthquake-resistant design of shallow foundations, the present study is concerned with the determination of the seismic bearing capacity of a circular footing resting on the surface of a heterogeneous purely cohesive semi-infinite soil layer. In the first part of the paper, a database, containing case histories of civil engineering structures that sustained a foundation seismic bearing capacity failure, is briefly presented, aiming at a better understanding of the studied phenomenon and offering a number of case studies useful for validation of theoretical computations. In the second part of the paper, the aforementioned problem is addressed using the kinematic approach of the Yield Design theory, thus establishing optimal upper bounds for the ultimate seismic loads supported by the soil-footing system. The results lead to the establishment of some very simple guidelines that extend the existing formulae for the seismic bearing capacity contained in the European norms (proposed for strip footings on homogeneous soils) to the case of circular footings and to that of heterogeneous cohesive soils.

Mots clés: fondations circulaires, capacité portante sismique, calcul à la rupture.

## 1 INTRODUCTION

Les observations sur site après les séismes importants des dernières décennies ont conduit à la formulation d'une méthodologie systématique pour la conception parasismique des fondations, comportant, en général, deux étapes principales. D'une part, l'évaluation de la résistance à la liquéfaction affecte principalement le sol de fondation et peut conduire le système fondation/superstructure à un état limite ultime (dans le sens des Eurocodes: rupture d'éléments structuraux, déplacements et/ou rotations excessifs, perte de fonctionnalité de la structure *etc*.). D'autre part, l'évaluation de la capacité portante sismique des fondations, phénomène mis en évidence après le séisme de Guerrero – Michoacán (Mexique, 1985) (Mendoza & Auvinet, 1988), où un nombre important de ruptures de fondations ont eu lieu accompagnées de grands déplacements verticaux, de grandes rotations permanentes et le développement d'un mécanisme de rupture dans le sol sans l'apparition de liquéfaction (*cf*. Figure 1).

La conception d'un système de fondation doit, dans un premier lieu, garantir que le risque de liquéfaction est éliminé. Ensuite elle doit choisir les paramètres de résistance appropriés et évaluer la capacité portante sismique du système en assurant un coefficient de sécurité suffisant. Cette procédure peut être inscrite dans un cadre de conception basée sur l'évaluation des déplacements (« displacement-based » design) et ultérieurement sur un nombre de critères de performance pour la structure (« performance-based » design). Dans tous les cas, la détermination du risque de liquéfaction et la connaissance de la capacité portante sismique demeurent deux éléments essentiels de la procédure de conception.

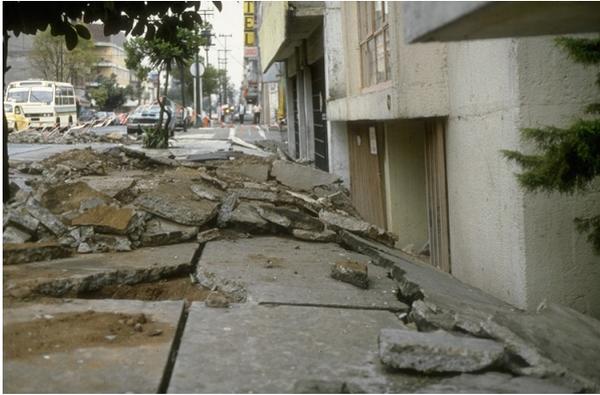

Figure 1. Rupture par perte de capacité portante après le séisme de Guerrero – Michoacán (Mexique, 1985).

Cette étude est consacrée à la deuxième étape de la conception parasismique de fondations : l'évaluation de la capacité portante sismique. À la différence du problème de liquéfaction, dont l'étude a été développée après les ruptures classiques observées lors du séisme de Niigata (Japon, 1964) (Seed & Idriss, 1967), la question de la capacité portante sismique est restée relativement sous-estimée et assez peu comprise. Pour cette raison, on a construit une base de données de structures et d'ouvrages de génie civil qui ont subi une rupture par perte de capacité portante lors d'un séisme récent, dans le double but : de contribuer à une meilleure compréhension du phénomène, mis en évidence par des observations de structures sur site et, en parallèle, de fournir des cas d'études réels, nécessaires pour la validation des méthodologies théoriques de calcul de la capacité portante sismique et de l'évaluation des déplacements permanents de la fondation.

Après une rapide présentation de la base de données, le traitement théorique du problème de la capacité portante d'une semelle circulaire reposant sur un sol purement cohérent hétérogène est décrit, ainsi que les conclusions les plus intéressantes du point de vue pratique.

## 2 BASE DE DONNÉES DE RUPTURES PAR PERTE DE CAPACITÉ PORTANTE

Pour inclure une rupture particulière dans la base de données, il fallait que celle-ci fût due à un dépassement de la capacité portante du système sol-fondation. Dans un premier lieu, les ruptures dues à la liquéfaction ont été exclues ainsi que les ruptures dues à une défaillance des éléments structuraux de la fondation.

### 2.1 *Sources*

Les sources pour la création de la base de données ont été diverses et comportaient principalement :
   *a*. Les articles dans des journaux scientifiques et dans des comptes rendus de congrès internationaux.
   *b*. Les volumes spéciaux de différents journaux consacrés à un grand séisme récent particulièrement dommageable, dont les volumes des *Earthquake Spectra* pour les grands séismes des vingt dernières années, les volumes spéciaux de *Soils and Foundations, etc.*
   *c*. Les informations disponibles en ligne sur un nombre de sites consacrés à la prévention des séismes et à la recherche en génie parasismique, dont l'Earthquake Engineering Research Institute (EERI) *etc*.
   *d*. Les rapports de reconnaissance après de grands séismes issus des Associations pour le Génie Parasismique, dont l'Association Française du Génie Parasismique (AFPS) *etc*.

Une liste complète des sources d'informations utilisées peut être trouvée dans Chatzigogos (2007).

### 2.2 *Structure de la base de données*

La base de données, qui a été créée, est constituée de fiches de données. Chaque fiche correspond à une structure particulière, qui a subi une rupture par perte de capacité portante. Presque 200 ouvrages ont été rassemblés couvrant un large spectre de combinaisons de types de structure, configurations de la fondation et conditions du sol (profil géotechnique). Une première classification a été faite par rapport au type de structure, en distinguant les catégories suivantes :
   *a*. Bâtiments.
   *b*. Ponts.
   *c*. Réservoirs – Infrastructure industrielle.
   *d*. Barrages.
   *e*. Structures portuaires.

Chaque fiche de données est composée de trois domaines de données décrivant la rupture : le premier comporte les données du séisme : *i*) le nom du séisme, *ii*) le lieu du séisme (pays – ville), *iii*) la date de l'événement sismique, *iv*) sa magnitude et *v*) les sources d'informations utilisées.

Le deuxième domaine de données est consacré à la description de la structure touchée, sa fondation et les conditions du sol de fondation. Plus précisément les données suivantes sont inclues : *i*) type de structure, *ii*) description de la structure, *iii*) type de fondation - description, *iv*) dimensions de fondation, *v*) sol de fondation.

Le troisième domaine de données concerne la description de la rupture elle-même et fournit les informations suivantes : *i*) déplacement permanent après le séisme, *ii*) rotation permanente après le séisme et *iii*) description de la rupture.

Toutes les autres informations disponibles sont aussi attachées à chaque fiche. Celles-ci peuvent être constituées des données de reconnaissance géotechnique du site, des dessins techniques de la structure, des photographies de la structure avant/après le séisme, *etc.*

Un exemplaire de fiche de données est présenté dans l'Annexe 1.

## 2.3 *Conclusions tirées de l'examen de la base de données*

L'examen des données reportées dans la base a conduit aux conclusions suivantes :

*Par rapport à l'identification des ruptures,* il a été reconnu que les ruptures par perte de capacité portante sont, dans la majorité des cas, fortement liées au phénomène de liquéfaction, à tel point qu'il est en général difficile d'isoler les deux phénomènes. Mis à part le cas extrême d'une liquéfaction générale touchant une grande région et entraînant la rupture des structures qui y sont localisées (et qui préservent souvent leur intégrité structurale), la plupart des ruptures observées sur site sont dues à une combinaison/interaction de deux phénomènes :
- augmentation des actions sismiques sur la fondation,
- affaiblissement de la résistance du sol de fondation à cause de la liquéfaction ou de l'effet cyclique du chargement.
De plus, la plupart du temps, la difficulté d'accès aux données et le manque d'informations ne permettent pas de tirer des conclusions certaines.

*Par rapport au type de structures concernées,* les conséquences les plus graves et les ruptures les plus spectaculaires ont été observées pour les ponts. Les déplacements et les rotations permanents induits au niveau de la fondation combinés avec les grandes dimensions de la structure, le système structurel habituel (appuis simples) et les faibles propriétés du sol de fondation (dépôts d'origine fluviatile) peuvent conduire à un endommagement grave de la superstructure.

*Par rapport aux propriétés du sol de fondation,* les ruptures par perte de capacité portante sont rencontrées surtout dans des sols cohérents mous, où le risque de liquéfaction ne se pose pas mais où la résistance cyclique du sol est insuffisante pour la reprise des actions sismiques. Un cas particulièrement intéressant est celui où la fréquence principale de la superstructure coïncide avec la fréquence de la couche de sol : la résonance lors d'une excitation sismique peut conduire à une augmentation considérable des efforts sur la fondation et éventuellement à une rupture par perte de capacité portante (Romo & Auvinet, 1991).

*Par rapport aux systèmes de fondation sujets à rupture,* il a été clairement mis en évidence que les systèmes conçus avec un coefficient de sécurité faible vis à vis des charges permanentes (FS<2) sont les plus affectés par une rupture par perte de capacité portante. Cela a été observé pour tous les types de fondations et particulièrement pour les fondations superficielles. Ce résultat a été vérifié ensuite tant expérimentalement que théoriquement.

L'observation des ruptures par perte de capacité portante a clairement montré que le contrôle des déplacements/rotations permanents, induits par le séisme, est le chemin à suivre pour la protection parasismique des structures, particulièrement en ce qui concerne les ruptures d'origine géotechnique. Curieusement, l'élément essentiel pour une conception basée sur les déplacements demeure l'évaluation de la capacité portante du système, puisque les déplacements/rotations permanentes ne sont rien d'autre que les déplacements accumulés lors de la sollicitation sismique, dès qu'il a y dépassement de la portance du système.

## 3 CAPACITÉ PORTANTE SISMIQUE D'UNE SEMELLE CIRCULAIRE

Ce travail se place dans le droit fil des travaux de Pecker & Salençon (1991), Salençon & Pecker (1995a, 1995b) et de Dormieux & Pecker (1995) et Paolucci & Pecker (1997), qui ont étudié la capacité portante sismique des semelles filantes sur des sols homogènes, soit purement cohérents, soit purement frottants en utilisant l'approche statique et surtout l'approche cinématique du Calcul à la Rupture (Salençon, 2002). L'élément novateur de l'étude présentée ici est le traitement d'une semelle circulaire (problème tridimensionnel) et l'introduction de l'hétérogénéité du sol.

### 3.1 *Formulation du problème*

Le problème consiste à déterminer les valeurs maximales de plusieurs paramètres de chargement, étant donnés trois éléments : *i*) la géométrie du système de fondation, *ii*) la résistance du sol et de l'interface sol – semelle et *iii*) le processus de chargement du système.

*Géométrie.* Il s'agit d'une semelle circulaire de rayon $r$ qui repose à la surface du sol, celui-ci décrit comme un demi espace.

*Critères de résistance.* Le sol est considéré isotrope et purement cohérent, et obéit au critère de Tresca. Sa cohésion varie linéairement avec la profondeur:

$$c = C_0 + Gz \qquad (1)$$

où $C_0$ désigne la cohésion du sol à la surface et $G$ le gradient vertical positif de cohésion. Deux cas ont été examinés : *i*) sol obéissant au critère de Tresca classique et *ii*) sol sans résistance à la traction : critère de Tresca tronqué en traction. L'interface a été considérée purement cohérente sans résistance à la traction (permettant ainsi la création d'une zone de décollement entre le sol et la semelle). Sa cohésion a été considérée égale à la cohésion du sol à la surface $C_0$.

*Processus de chargement.* La semelle est soumise à l'action d'une force verticale $N$ modélisant le poids de la superstructure et d'une force horizontale $V$ et d'un moment $M$ décrivant l'excitation inertielle de la superstructure lors du séisme. De plus, des forces d'inertie volumiques de direction horizontale $F_h$ sont considérées dans le sol. Celles-ci modélisent les forces inertielles lors de la propagation des ondes sismiques à travers le volume de sol. Le vecteur $Q$ des paramètres de chargement du système est alors formulé comme ci dessous :

$$\underline{Q} = \left(N, V, M, F_h\right) \qquad (2)$$

Dans le cas d'une structure à un degré de liberté : $M = Vh$, $h$ : hauteur de la structure . Les données du problème sont présentées sur la Figure 2.

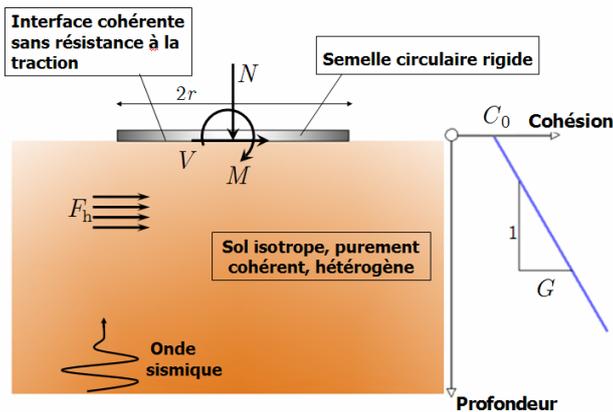

Figure 2. Capacité portante sismique d'une semelle circulaire reposant à la surface d'un sol purement cohérent hétérogène.

On note que l'utilisation du critère de Tresca pour le sol implique que le poids volumique du sol $\gamma$ et les forces d'inertie verticales dans le sol sont sans influence sur les charges ultimes supportées par le système (*cf.* Salençon & Pecker, 1995a, b).

*Hypothèses supplémentaires.* Pour faciliter le traitement du problème, deux hypothèses supplémentaires sont introduites : A) les forces d'inertie $F_h$ et la force horizontale $V$ sur la semelle sont considérées colinéaires et le moment $M$ est perpendiculaire à leur direction. Le fondement physique de cette hypothèse revient à considérer que les efforts $V, M$ sur la semelle proviennent de l'excitation inertielle d'une structure à un degré de liberté. B) Les forces d'inertie sont uniformes dans l'espace. Cette seconde hypothèse est d'autant plus réaliste que la dimension de la semelle est petite par rapport à celle de la couche de sol, dans laquelle se propagent les ondes sismiques.

### 3.2 Résolution du problème

Le problème est traité par la méthode cinématique du Calcul à la Rupture. Les détails pour l'application de cette méthode peuvent être trouvées dans Salençon (2002). La méthode fournit une approximation par excès (bornes supérieures) pour les charges ultimes supportées par le système en examinant une série de champs de vitesse virtuelle cinématiquement admissibles. Pour le problème traité, un champ de vitesse est cinématiquement admissible s'il respecte la rigidité parfaite de la semelle et est nul à l'infini.

*Construction des champs de vitesse virtuelle.* Au premier abord, on note que, puisque les forces d'inertie sont considérées uniformes dans l'espace et que le domaine du sol est considéré comme semi infini, les forces d'inertie maximales supportées, les autres paramètres de chargement étant nuls, sont en fait nulles. Ce résultat n'a évidemment aucune vraisemblance physique. En effet, la forme des champs de vitesse lors d'une rupture par perte de capacité portante a été expérimentalement étudiée par Knappett *et al.* (2006), qui ont effectué des essais sur la table vibrante de l'Université de Cambridge (RU). Ils ont pu identifier le mécanisme de rupture dans le sol en utilisant la technique de mesure des déplacements (Particle Image Velocimetry). Un résultat typique est présenté sur la Figure 3.

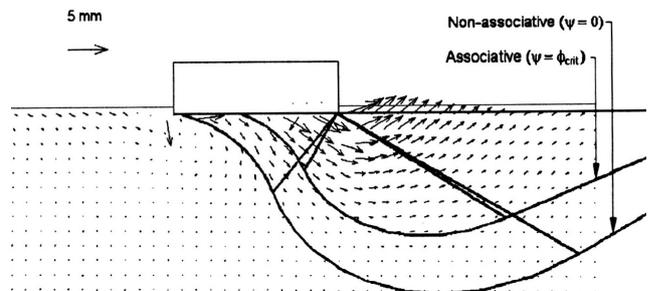

Figure 3. Identification de mécanismes de rupture dans des essais sur table vibrante (d'après Knappett *et al.* 2006).

Les résultats expérimentaux de Knappett *et al.* (2006) permettent de surmonter la difficulté présentée par le problème des forces d'inertie nulles en ne considérant que des mécanismes de rupture (champs de vitesse virtuelle) correspondant aux observations expérimentales et sur site.

Ainsi, on *anticipe* la forme du mécanisme de rupture en examinant les trois familles de champs de vitesse virtuelle suivants :

*a. Mécanismes translationnels.* Le mécanisme de rupture est créé dans le sol par une translation pure de la semelle.

*b. Mécanismes purement rotationnels.* Le mécanisme de rupture est créé dans le sol par une rotation de la semelle autour d'un axe de rotation qui se trouve au dessus de la semelle.

*c. Mécanismes rotationnels avec cisaillement.* Le mécanisme de rupture est créé dans le sol par une rotation de la semelle autour d'un axe de rotation qui se trouve au dessous de la semelle. Une zone de cisaillement est créée dans le sol, d'où le nom des mécanismes.

La forme des mécanismes dans chaque famille est contrôlée par un nombre de paramètres géométriques. En changeant la valeur de ces paramètres on obtient une infinité de champs de vitesse virtuelle. Une procédure d'optimisation permet de déterminer le champ de vitesse (en identifiant la famille et la valeur des paramètres géométriques), qui va fournir les bornes supérieures minimales pour chaque combinaison des charges ultimes du système. Tous les détails concernant les champs de vitesse virtuelle considérés sont donnés dans Chatzigogos (2007).

### 3.3 Résultats

Les résultats sont présentés comme des surfaces ultimes tracées dans l'espace des paramètres de chargement $(N,V,M)$ pour différentes valeurs de $F_h$ et du paramètre sans dimension $k = rG/C_0$.

*Erreur induite par les bornes supérieures.* Les résultats établis ne sont pas conservatifs : il est donc essentiel d'avoir une idée de l'erreur induite par les bornes supérieures. Une comparaison des résultats peut être effectuée avec les solutions exactes de Salençon & Matar (1982) pour le cas axisymétrique (la semelle est sollicitée uniquement par une force verticale $N$, les autres paramètres étant nuls). Les résultats pour un sol sans résistance à la traction sont présentés dans le Tableau 1. Il s'agit de la force verticale $N$ supportée par la semelle, normalisée par l'aire de la semelle et $C_0$. La comparaison montre que l'erreur augmente de 10% pour un sol homogène jusqu'à presque 25% pour un sol fortement hétérogène. Pour les cas de chargements sismiques habituels, l'erreur sera vraisemblablement inférieure aux valeurs présentées dans le Tableau 1, puisque les mécanismes de rupture considérés sont plus adaptés à la description des chargements avec $M,V,F_h$ élevées qu'au cas axisymétrique.

*Modification de la formule analytique de l'Eurocode 8.* La formule analytique proposée (Eurocode 8) décrit la surface ultime de la fondation dans l'espace $(N, V, M)$ et concerne uniquement les semelles filantes sur des sols homogènes purement cohérents et purement frottants. Pour les sols cohérents, il s'agit de l'équation suivante :

$$\frac{\left(1-e\overline{F_h}\right)^{c_V'}\left(\beta\overline{V}\right)^{c_V}}{\left(\overline{N}\right)^a\left[\left(1-m\overline{F_h}^k\right)^{k'}-\overline{N}\right]^b} + \frac{\left(1-f\overline{F_h}\right)^{c_M'}\left(\gamma\overline{M}\right)^{c_M}}{\left(\overline{N}\right)^c\left[\left(1-m\overline{F_h}^k\right)^{k'}-\overline{N}\right]^d} - 1 \leq 0 \quad (3)$$

soumise aux contraintes :

$$0 < \overline{N} < \left(1-m\overline{F_h}^k\right)^{k'}, \quad |\overline{V}| \leq 1/(\pi+2) \quad (4)$$

où on définit les quantités suivantes : $\overline{N} = N/N_{\max}$, $\overline{V} = V/N_{\max}$, $\overline{M} = M/BN_{\max}$, $\overline{F_h} = \rho a_h B/C_0$, $N_{\max} = (\pi+2)BC_0$ (capacité portante statique), $B$ : largeur de la semelle, $\rho$ : masse volumique du sol, $a_h$ : accélération sismique horizontale. Les autres paramètres dans (3) sont des constantes.

Tableau 1. Comparaison des bornes supérieures et des solutions exactes pour le cas du chargement axisymétrique.

| $\dfrac{Gr}{C_0}$ | Valeurs Exactes (Salençon & Matar, 1982) | Bornes Supérieures | Erreur (%) |
|---|---|---|---|
| 0.00 | 6.065 | 6.71 | 10.63% |
| 0.50 | 6.933 | 7.79 | 12.36% |
| 1.00 | 7.614 | 8.61 | 13.08% |
| 3.00 | 10.080 | 12.15 | 20.54% |
| 5.00 | 11.724 | 14.95 | 27.52% |

Les modifications suivantes sont proposées pour permettre l'application de l'équation (3) dans le cas de semelles circulaires et dans le cas de sols hétérogènes.

*Semelles circulaires – sols homogènes.* Il suffit de remplacer $B$ par le diamètre $D$ de la semelle et la valeur de la capacité portante statique $N_{\max}$ par la valeur équivalente pour les semelles circulaires donnée dans le Tableau 1 ($k=0$) : $N_{\max} = 6.06\pi D^2 C_0$.

*Sols hétérogènes.* Dans le cas d'un sol cohérent avec un gradient de cohésion non nul, il faut prendre en compte l'action favorable du gradient de cohésion qui diminue l'effet négatif des forces d'inertie. Ainsi, on introduit des forces d'inertie fictives dans (3) qui sont données par :

$$\overline{F_h}' = \zeta \overline{F_h}, \quad \zeta = \frac{N_{\max}^{(0)}}{N_{\max}^{(k)}} \quad (5)$$

où $N_{\max}^{(0)}$ est la capacité portante du sol obtenue pour $G=0$ et $N_{\max}^{(k)}$ la capacité portante du sol hétérogène. Les autres paramètres de chargement sont donnés par : $\overline{N} = N/N_{\max}^{(k)}$, $\overline{V} = V/N_{\max}^{(0)}$, $\overline{M} = M/DN_{\max}^{(k)}$. La même modification peut être introduite aussi pour les semelles filantes sur sols hétérogènes. La valeur de $N_{\max}^{(k)}$ est donnée par Salençon & Matar (1982) tant pour les semelles circulaires (*cf*. Tableau 1) que pour les semelles filantes.

Les bornes supérieures établies sont comparées sur la Figure 4 avec les résultats issus de l'application de l'équation de l'Eurocode 8, modifiée comme indiqué ci-dessus. Il s'agit du diagramme d'interaction $M - V$ pour $N = 1/3 N_{max}^{(k)}$ et pour : a) $k = 0$ et b) $k = 1$.

La comparaison montre que les deux familles de courbes sont en très bon accord notamment dans la partie : $M > 0, V > 0$ ; les modifications simplificatrices proposées peuvent être considérées acceptables.

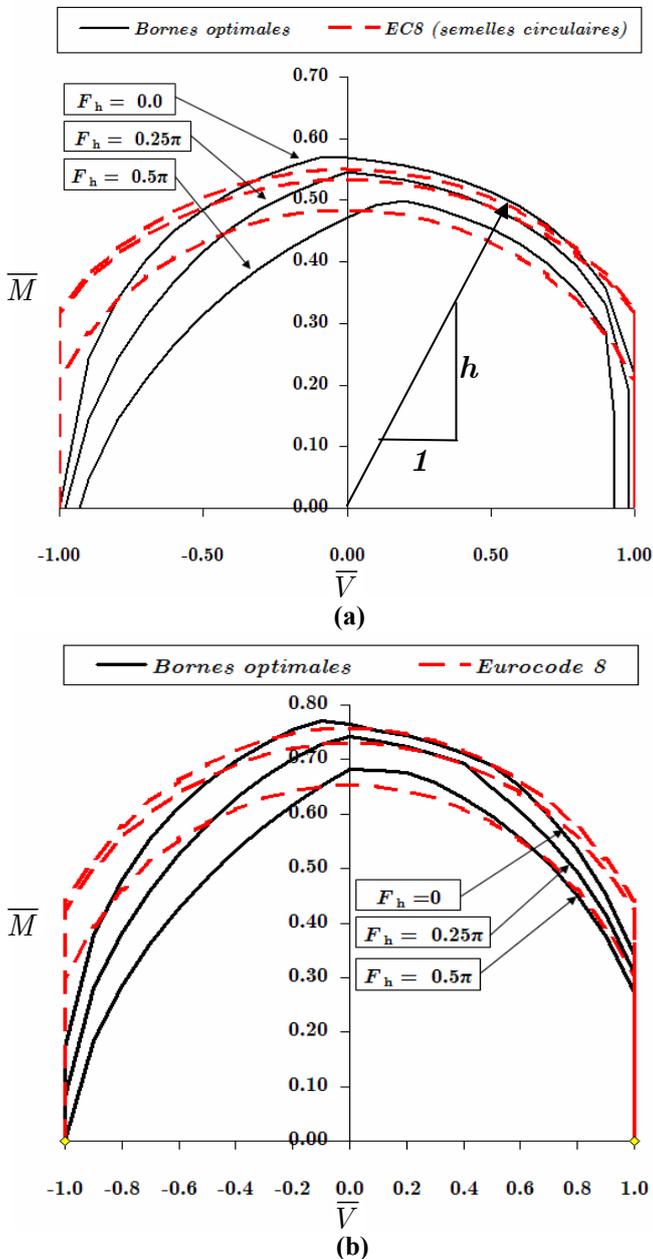

Figure 4. Comparaisons des bornes supérieures et de l'équation modifiée de l'Eurocode 8. Diagramme d'interaction $\overline{M} - \overline{V}, \overline{N} = 1/3 N_{max}^{(k)}$ : a) $k = 0$, b) $k = 1$

## 4 CONCLUSIONS

Dans cette étude, on a traité le problème de la capacité portante sismique d'une semelle circulaire sur un sol cohérent hétérogène. La base de données présentée a montré que la perte de capacité portante lors d'un séisme est un problème réel, même s'il est souvent difficile de le dissocier du phénomène de liquéfaction. Le traitement du problème par l'approche cinématique du Calcul à la Rupture a permis l'établissement des bornes supérieures optimales, satisfaisantes pour la conception, ainsi que l'extension de la formule de l'Eurocode 8 au cas des semelles circulaires et des sols cohérents hétérogènes avec des modifications mineures.

REMERCIEMENTS

Le premier auteur remercie l'École Polytechnique et la Fondation « Alexandros S. Onassis » pour le support financier pendant l'exécution de cette étude.

ANNEXE 1

Seismic Damage to Foundations of Structures

# DATASHEET FOR SEISMIC FOUNDATION DAMAGE

## EARTHQUAKE DATA

**Earthquake:** GUERRERO - MICHOACAN
**Location:** Guerrero - Michoacan
**Country:** Mexico
**Date:** 9/19/1985
**Magnitude:** M8.0
**Reports Comments:** Earthquake Spectra, Volume 4, Issue 3 - 4, Vol. 5, Issue 1 - AFPS Mission - EEQIS Library

## STRUCTURE DATA

**Type of Structure:** Building
**Description:** IRREGULAR FRICTION PILE BUILDING
**Foundation Type:** Foundation on rigid box (144KPa) and friction piles
**Foundation Dimensions:** Irregular area of 160 square m.
**Soil Conditions:** lacustrine soft clay of Mexico city

## FAILURE DESCRIPTION

**Displacement:** -
**Rotation:** Total rotational collapse
**Description:** The irregular shape of the building in plan led to the existance of an oblique axis with minimum capacity to resist overturning moments. Eccentricity=1.4m. Total collapse by overturning in NW. FS,static=1.7. FS,seismic=1. It can be considered that the short piles of this foundation were working at their limit capacity under static conditions, and that a significant contact pressure existed at the slab level. The increasing plastic deformations of the soil induced by the seismic cyclic stresses led in turn to higher overturning moments at the base of this slender structure by p-δ effect, until the bearing capacity of the foundation slab was overcome; the contribution to the overturning capacity of the lateral reactions on the walls of the substructure was probably negligible due to its shallow depth. It is important to note that it was not necessary to include in the analysis any consideration about a possible degradation of the adherence between piles and soil under cyclic loading to explain the failure. The importance of adherence degradation was probably overemphasized in some evaluations published after the earthquake.

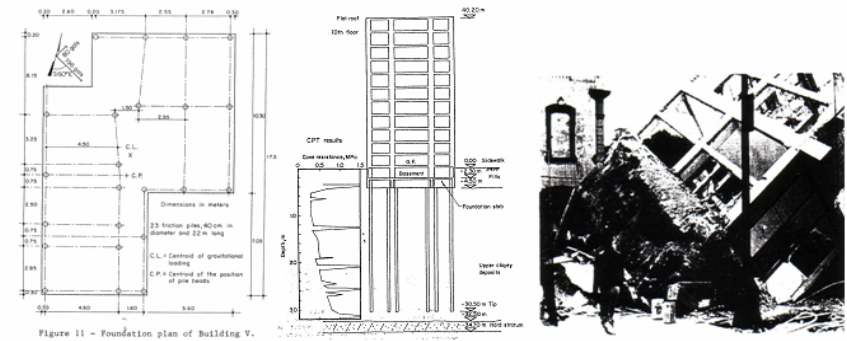

Figure 11 – Foundation plan of Building V.

Photo Credits: Mendoza & Auvinet (1988)

Charisis Th. Chatzigogos

LMS – École Doctorale De l'École Polytechnique